\documentclass{iau}
\usepackage{graphicx} 

\title[IAUS291.~~PSR~B1259-63 spectrum evolution: detailed study] 
{Binary pulsar B1259-63 spectrum evolution: detailed study} 

\author[M. Dembska, J. Kijak \& W. Lewandowski]  
{Marta Dembska, Jaros\l{}aw Kijak \and  Wojciech Lewadowski}

\affiliation{Kepler Institute of Astronomy, University of Zielona G\'ora\\
Lubuska 2, 65-265 Zielona G\'ora, Poland \\
 email: {\tt marta@astro.ia.uz.zgora.pl}}

\pubyear{2012}
\volume{291}  
\jname{\mbox{Neutron Stars and Pulsars: Challenges and Opportunities after 80 years}}
\editors{J. van Leeuwen, ed.} 
\begin{document}

\maketitle

\begin{abstract}
We studied the radio spectrum of PSR~B1259-63 in an unique binary with Be star LS~2883 and showed that the shape of the spectrum depends on the orbital phase. We proposed a qualitative model which explains this evolution. We considered two mechanisms that might influence the observed radio emission: free-free absorption and cyclotron resonance. Recently published results have revealed a new aspect in pulsar radio spectra. There were found objects with turnover at high frequencies in spectra, called gigahertz-peaked spectra (GPS) pulsars. Most of them adjoin such interesting environments as HII regions or compact pulsar wind nebulae (PWN). Thus, it is suggested that the turnover phenomenon is associated with the environment than being related intrinsically to the radio emission mechanism. Having noticed the apparent resemblance between the B1259-63 spectrum and the GPS, we suggest that the same mechanisms should be responsible for both cases. Therefore, the case of B1259-63 can be treated as a key factor to explain the GPS phenomenon observed for the solitary pulsars with interesting environments and also another types of spectra (e.g. with break).

\keywords{pulsars: general, individual (B1259-63) - stars: winds, outflows - ISM: general, magnetic fields - radiations mechanism: non-thermal}
\end{abstract}

\firstsection 
\section{The~gigahertz-peaked spectra pulsars}
Generally, the~observed radio spectra of~most pulsars can be modelled as a~power law with~negative spectral indices of~about~-1.8 (\cite{Maron2000}). If a~pulsar can be observed at~frequencies low enough (i.e.~\mbox{100-600~MHz}), it may also show a~low-frequency turnover in~its spectrum (\cite{Sieber1973}; \cite{Malofeev1994}). On the~other hand, Lorimer \etal\ (1995) mentioned three pulsars which have positive spectral indices in~the~frequency range 300-1600~MHz. Later, Maron~\etal\ (2000) re-examined spectra of~these pulsars taking into account the~data obtained at~higher frequencies (above~1.6~GHz) and~consequently were the~first to demonstrate a~possible existence of~spectra with~turnover at~high frequencies, about 1~GHz. Kijak~\etal\ (2011a) provided a~definite evidence for~a~new type of~pulsar radio spectra. These spectra show the~maximum flux above 1~GHz, while at~higher frequencies the~spectra look like a~typical pulsar spectrum. At lower frequencies (below 1~GHz), the~observed flux decreases, showing a positive spectral index (\cite{Kijak2011a}). They called these objects the~gigahertz-peaked spectra (GPS) pulsars. A~frequency at~which such a~spectrum shows the~maximum flux was called the~peak frequency. Kijak~et~al.~(2011a) also indicated that~the~GPS pulsars are relatively young objects, and~they usually adjoin such interesting environments as HII regions or compact pulsar wind nebulae. Additionally, some of~them seem to~be coincident with~the~known but sometimes unidentified X-ray sources from third EGRET Catalogue or HESS observations. We can assume that~the~GPS appearance owes to the environmental conditions around the~neutron stars rather than to the~radio emission mechanism.
\section{PSR~B1259-63 spectrum evolution}
PSR~B1259-63 was also listed by~Lorimer~\etal\ (1995) as a pulsar with~positive spectral index. Therefore, it seems a natural candidate to~be classified as the~GPS pulsar. This pulsar is in an unique binary with a~massive main-sequence Be star. \mbox{PSR~B1259-63} has a~short period of~48 ms and~a~characteristic age of~330~kyr. Its average dispersion measure~(DM) is~about 147~pc~cm$\mathrm{^{-3}}$ and~the~corresponding distance is about 2.75 kpc. The companion star LS~2883 is a~10-mag massive Be star with~a~mass of about 10M$\mathrm{_{\odot}}$  and~a~radius of~6R$\mathrm{_{\odot}}$. Be stars are generally believed to~have  a~hot tenuous polar wind and~a~cooler high-density equatorial disc. The~PSR~B1259-63/LS~2883 emits unpulsed non-thermal emission over a~wide range of~frequencies ranging between radio and~$\mathrm{\gamma-}$rays, and~its flux varies with~orbital phase. 
\begin{figure}[!hb]
 \begin{tabular}{cc}
   \includegraphics[width=0.49\textwidth]{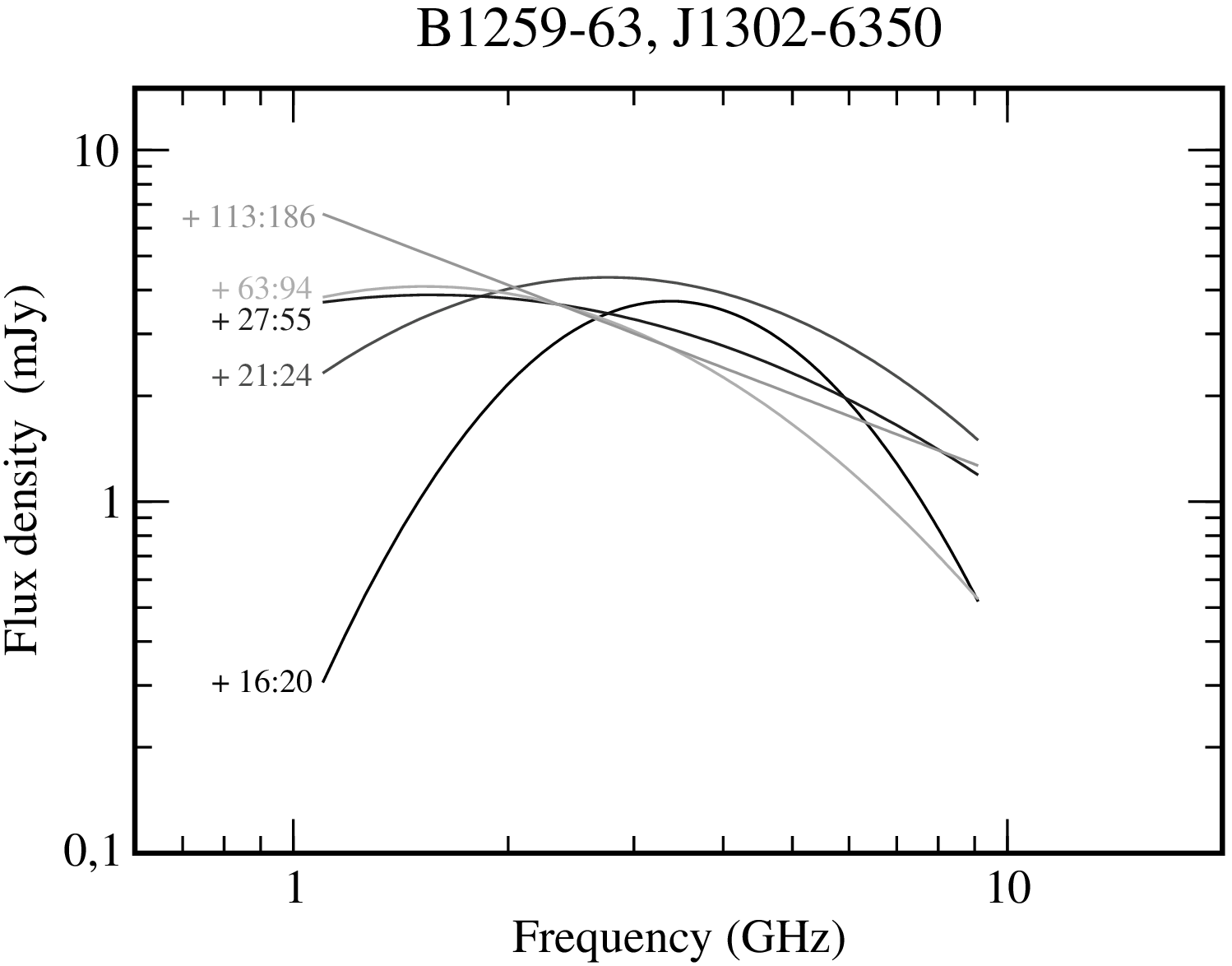}
   \includegraphics[width=0.49\textwidth]{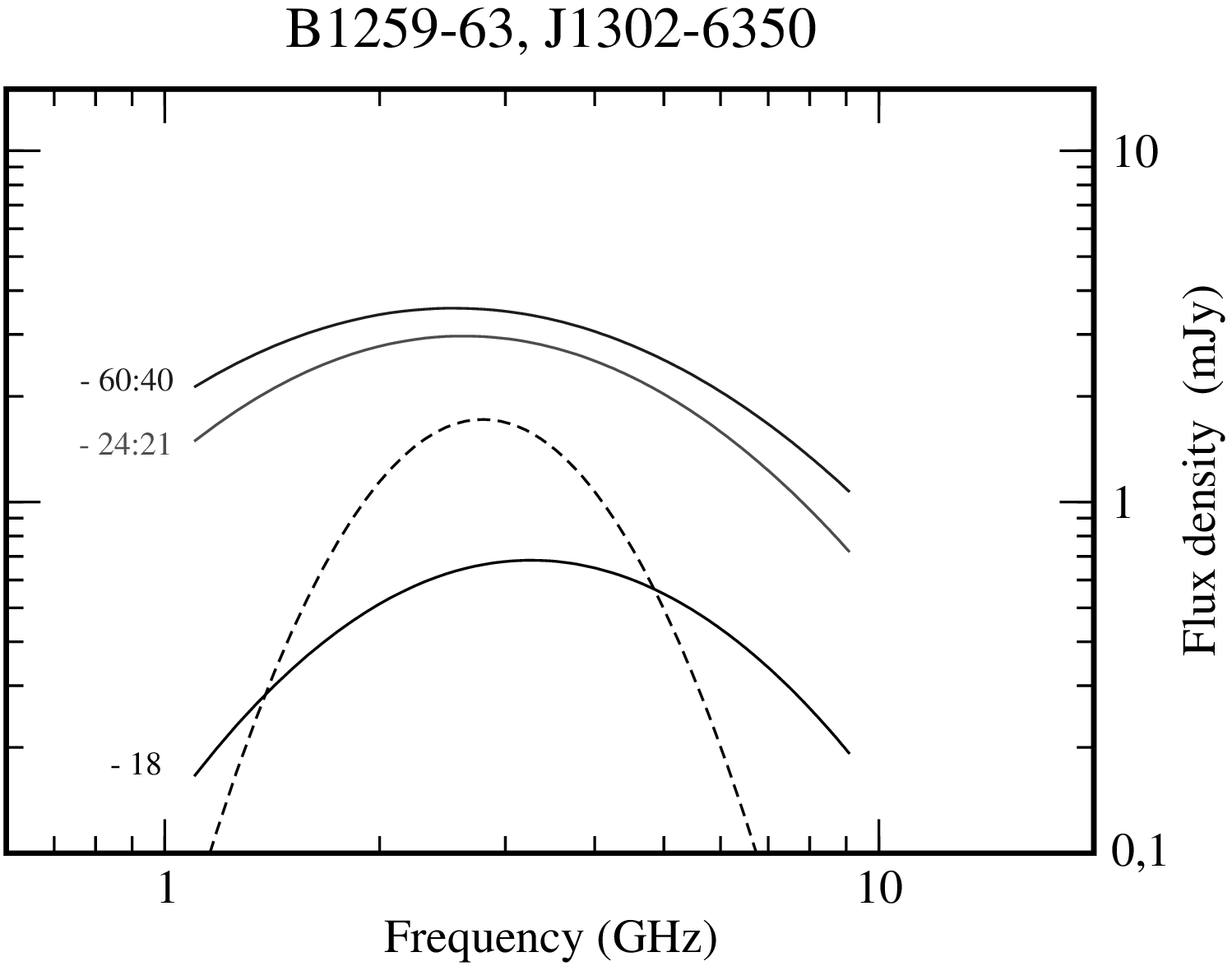}\\
 \end{tabular}
   \caption{The fits to the B1259-63 spectra for the each orbital phase range (\cite{Kijak2011b}), from 60 d prior to periastron (left panel) up to 186 d after it (right panel).}
   \label{evolution}
\end{figure}
We~studied the~radio spectrum of~B1259-63 (\cite{Kijak2011b}). We~analysed the~available  measurements of~the~pulsed flux obtained during three periastron passages (1997, 2000 and~2004). Our analysis showed that~this pulsar undergoes a~spectrum evolution due to~orbital motion (see~Fig.\ref{evolution}). We suggested that~this effect is caused by~the~interaction of the~radio waves with~the~Be star environment. While the~eclipse itself can be naturally explained by~free-free absorption in~the~stellar disc, the~disc alone is not enough to~explain the~spectra evolution. In~addition, we have shown that~the~peak frequency also~depends on~the~orbital phase and therefore it varies with the changes of the pulsar environment. We argued that~such behaviour can be explained by~the~radio-wave absorption in~the~magnetic field associated with~the~disc. We proposed a~qualitative model  which explains this evolution. We argued that~the~observed variation of~the~spectra is  caused~by~a~combination of~two effects: the~free-free absorption in~the~stellar wind and~the~cyclotron resonance in~the~magnetic field. This field is associated with the disc and is infused by the relativistic particles of the pulsar wind.

\textbf{B1259-63 as key factor to~explain the~GPS phenomenon.} Having noticed the~apparent resemblance between the~B1259-63 spectrum and~the~GPS, we suggested that~the~same mechanisms should be responsible for~both cases~(\cite{Kijak2011b}). Thus, we can conclude that the~GPS feature should be caused by some external factors rather than by the~emission mechanism. On~the other hand, the~GPS pulsars are isolated radio pulsars and~therefore, we cannot draw a direct analogy between the PSR~\mbox{B1259-63}/LS~2883 system and~the~GPS pulsars, as the~latter have no companion stars and/or discs. But the~GPS pulsars apparently are surrounded by some kind of~environment that can affect the~spectra of~those pulsars in the~same way as the~stellar wind affects the~B1259-63 spectrum. All GPS pulsars have relatively high DMs that, in some cases, are too large to be accounted for by the Galactic electron density, and thus, we can speculate that there is a quite high particle density in the vicinity of these  pulsars (see also \cite{Kijak2011a}). Thus, we believe that~this binary system can hold the~clue to~the~understanding of~gigahertz-peaked spectra of~isolated pulsars. The~only difference could be an invariable shape of~the~GPS.
\section{PSR B1259-63 spectrum evolution: detailed study}
Using the~same database (\cite{Johnston1999}, \cite{Connors2002}, \cite{Johnston2005}) we constructed spectra for~chosen observing days. We analysed the shapes of the~\mbox{B1259-63} spectrum at~various orbital phase ranges as it was done previously (\cite{Kijak2011b}) and~obtained results are consistent with~those for~intervals. 
\begin{figure}[!hc]
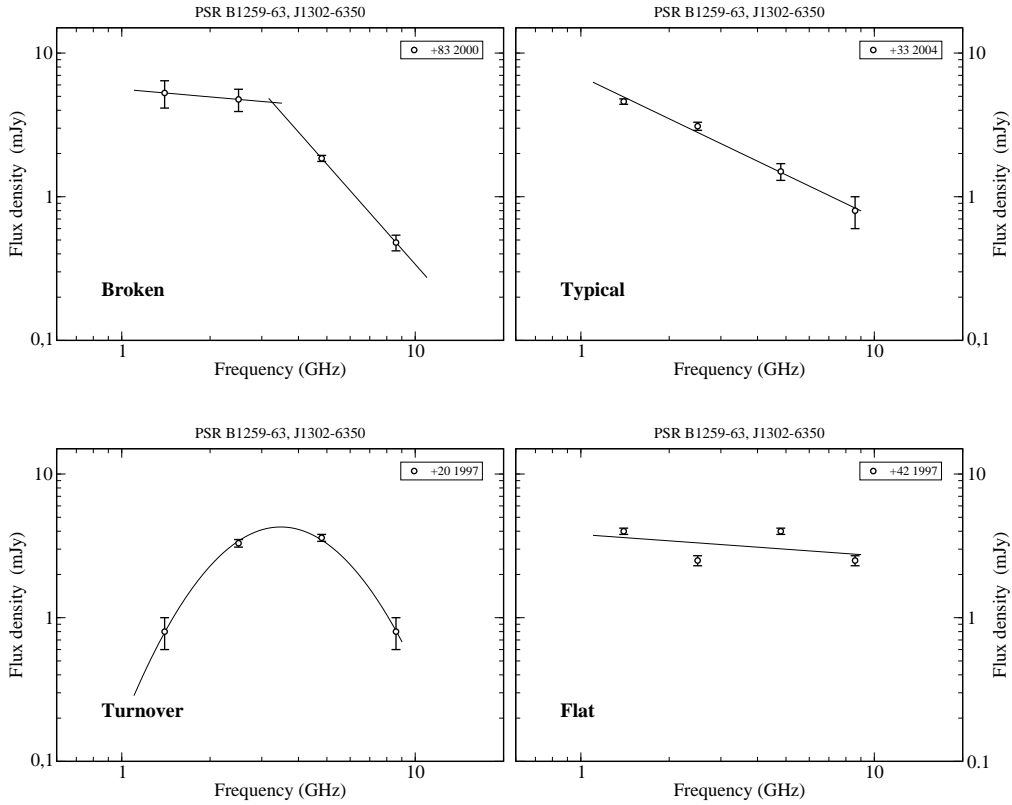

 \begin{tabular}{cc}
   \includegraphics[width=0.49\textwidth]{b1259-63_broken.eps}
   \includegraphics[width=0.49\textwidth]{b1259-63_typical.eps}\vspace{0.5cm}\\
   \includegraphics[width=0.49\textwidth]{b1259-63_turnover.eps}
   \includegraphics[width=0.49\textwidth]{b1259-63_flat.eps}\\
 \end{tabular}
 \caption{The fits to the B1259-63 spectra for chosen days. Each panel shows different type of spectra.}
 \label{detailed}
\end{figure}
The~flux at~the~given frequency apparently changes with~orbital phases. When the~pulsar is close to~periastron, the~flux generally decreases at~all observed frequencies and~the~most drastic decrease is observed at~the~lowest frequency. Moreover, we noticed all types of radio pulsar spectra.

PSR~B1259-63 is a object with relatively high dispersion measure which means that its transition frequency is very high. This implies that we definitely have to take into consideration both refractive (RISS) and diffractive (DISS) scintillations when analysing spectra for a given day. We used diffractive bandwidth $\Delta f_{\mathrm{DISS}}$ and timescale $\Delta t_{\mathrm{DISS}}$ from scintillation observations of the pulsar made far from periastron at 4.8 GHz and 8.4 GHz (\cite{scin98}) to estimate values of these parameters at 1.4 GHz and 2.4 GHz assuming $\Delta t_{\mathrm{DISS}}\propto f^{1.2}d^{-0.6}$, where $f$ and $d$ denote frequency and distance respectively.  We estimated the values of $\Delta t_{\mathrm{DISS}}$ to be ranging from 40~s at 1.4~GHz to 360~s at 8.4~GHz which suggests that diffractive scintillations should not affect the average flux measurements (observing sessions was usually 4 hours long). Roughly estimated refractive timescales vary from 12~hours at~8.4~GHz to more than 20~days at~1.4~GHz. However, for lower frequencies the modulation index is relatively small which means lower uncertainty estimates when measuring flux. High frequency observations will be affected by refractive scintillations what leads to conclusion that flux values should be averaged over epochs and/or orbital phase intervals to be
more reliable.

\section{Conclusions}
Close to~the~periastron point the spectra of~B1259-63 resemble those of~the GPS pulsars.  The spectrum for the~orbital epochs further from the~periastron point are more consistent  with~typical pulsar spectra (i.e.~power-law and~broken). Moreover, detailed study of~PSR~B1259-63 spectra revealed the~appearance of~all~types of~spectral shapes, including a~flat spectrum (see Fig.~\ref{detailed}).

We believe that the case of~B1259-63 can be treated as~a~key factor to~our understanding of~not only the~GPS phenomenon (observed for the solitary pulsars with~interesting environments) but also other types of~untypical spectra as~well (e.g.~flat or~broken spectra). This in turn would suggest, that the~appearance of~various non-standard spectra shapes in~the~general population of~pulsars can be caused by~peculiar environmental conditions.

\acknowledgments MD is a scholar within Sub-measure 8.2.2 Regional Innovation Strategies, Measure 8.2 Transfer of knowledge, Priority VIII Regional human resources for the economy Human Capital Operational Programme co-financed by European Social Fund and state budget.

\end{document}